\documentclass[manuscript]{aastex}

\usepackage[displaymath]{lineno}
\usepackage{color}

\slugcomment{Prepared as article for the American Naturalist. 

Contains appendices A and B,  3 tables and 7 figures (figures 2-7 are color figures)
}

\shorttitle{Effect of sex on an empirical landscape}
\shortauthors{de Visser et al.}

\begin{document}

\title{Exploring the effect of sex on an empirical fitness landscape}


\author{J. Arjan G. M. de Visser\altaffilmark{1,*}, Su-Chan Park\altaffilmark{2,\dag},  and Joachim Krug\altaffilmark{2,\ddag} }
\affil{
1. Laboratory of Genetics, Wageningen University, Arboretumlaan 4, 6703BD Wageningen, The Netherlands;
2. Institute for Theoretical Physics, University of Cologne, Z\"ulpicher Str. 77, 50937 Cologne, Germany
}
\altaffiltext{*}{Corresponding author; e-mail: Arjan.devisser@wur.nl}
\altaffiltext{\dag}{E-mail: psc@thp.uni-koeln.de}
\altaffiltext{\ddag}{E-mail: krug@thp.uni-koeln.de}

\linenumbers

\begin{abstract}
The nature of epistasis has important consequences for the evolutionary significance of sex and recombination.  Recent efforts to find negative epistasis as source of negative linkage disequilibrium and associated long-term sex advantage have yielded little support.  Sign epistasis, where the sign of the fitness effects of alleles varies across genetic backgrounds, is responsible for ruggedness of the fitness landscape with implications for the evolution of sex that have been largely unexplored.  Here, we describe fitness landscapes for two sets of strains of the asexual fungus \emph{Aspergillus niger} involving all combinations of five mutations.  We find that $\sim 30$\% of the single-mutation fitness effects are positive despite their negative effect in the wild-type strain, and that several local fitness maxima and minima are present.  We then compare adaptation of sexual and asexual populations on these empirical fitness landscapes using simulations.  The results show a general disadvantage of sex on these rugged landscapes, caused by the break down by recombination of genotypes escaping from local peaks.  Sex facilitates escape from a local peak only for some parameter values on one landscape, indicating its dependence on the landscape's topography.  We discuss possible reasons for the discrepancy between our results and the reports of faster adaptation of sexual populations.
\end{abstract}




\section{Introduction}
The way genes interact in their effect on a phenotype or fitness, called epistasis, has important implications for evolution, including the evolution of sex and recombination~\citep{Wetal2000}.  Epistasis entails any deviation from independent (i.e. additive or multiplicative, depending on the evolutionary model used) gene effects, and hence encompasses many different forms of gene interaction.  At least two forms of epistasis are relevant for the question why sex and recombination have evolved.  The first is a one-dimensional type of magnitude epistasis, called \emph{negative epistasis}, where the fitness effects of alleles all have the same sign (i.e. all are either deleterious or beneficial) and the fitness of genotypes with many alleles is lower than expected from the product of their individual effects~\citep{B1995,OL2002}. Negative epistasis provides a long-term advantage to sex and recombination by causing negative linkage disequilibria of alleles affecting fitness~\citep{B1995}.  By breaking down linkage disequilibria and increasing fitness variation, sex may facilitate the response to selection due to the net production of genotypes carrying extreme numbers of fitness alleles.  Besides negative epistasis, genetic drift combined with directional selection can cause negative linkage disequilibria~\citep{B1995,dVE2007,K1993,OL2002}. However, magnitude epistasis also has consequences for fitness in the next generation, and it is the balance between short- and long-term effects of sex that determines the conditions favoring selection for sex and recombination~\citep{B1995,OL2002}.  Most experimental studies have concentrated on detecting magnitude epistasis, but despite increasing efforts during the last decade the support for negative epistasis is limited~\citep{dVE2007,Ketal2007}.

The second type of epistasis with potential implications for the evolution of sex is \emph{sign epistasis}, where the sign of the fitness effects of alleles varies across genetic backgrounds~\citep{Wetal2005}.  Sign epistasis causes adaptive constraints by limiting the number of mutational pathways that can be taken by natural selection.  If sign epistasis is locally consistent, such that all mutational pathways leading to a particular genotype show the same fitness change upon approach of that genotype, these constraints are most severe and cause local fitness peaks and valleys~\citep{Wetal2005}.  

The effect of sex on rugged fitness landscapes has received little
attention, but is likely to depend on the landscape's topography, and
thus is largely an empirical problem.  Unlike tests of negative
epistasis, which are possible with phenotypic analyses (see
\citet{Ketal2007}), conclusive tests of sign epistasis require
knowledge of the genotypes involved.  Moreover, to study evolutionary
constraints from sign epistasis, a systematic and detailed description
of the fitness landscape is needed, requiring the construction and
fitness measurement of many genotypes.  These requirements clearly
constrain empirical studies of fitness landscapes.  However, with the
advent of genomic techniques recent studies have begun to tackle this
problem, which found ample support for sign
epistasis~(Weinreich et al. 2006; Weinreich et al. 2005; Poelwijk et al.
2007; Salverda et al. unpublished).
Indirect
support for fitness landscapes with multiple peaks has come from the
analyses of fitness trajectories of evolution experiments with
micro-organisms, where replicate populations sometimes approach
different fitness
maxima~\citep{Betal2003,BC2000,Ketal1994,Retal2008,Setal2007}. Outside
the realm of evolutionary computation~\citep{WW2006}, theoretical
studies of the effect of sex on rugged fitness landscapes are rare.
One study, which modeled a specific landscape with a single narrow
ridge of increasing fitness, found that sex slows down
adaptation~\citep{KK2001}, while a study that modified this model to
include isolated peaks occupied by a polymorphic population found a
sex benefit~\citep{WW2006}.  Finally, a study using a multi-locus
rugged fitness landscape generated according to Kauffman's 
NK-model ~\citep{Kauffman1993} found faster adaptation due to recombination for
the parameter values used, particularly when recombination rates were negatively associated with fitness~\citep{HB2003}.

The aim of the present paper is twofold.  We first analyze fitness data of the filamentous fungus \textit{Aspergillus niger} in order to detect and quantify the occurrence of sign epistasis and multiple fitness peaks, and then explore the effects of sex on these empirical landscapes using simulations.  The data involve two collections of strains that each carry all possible combinations of five phenotypic marker mutations with individually deleterious effect (four of which are shared by the two collections).  These data were collected and analyzed previously to test for prevailing negative epistasis among deleterious mutations, for which we found no significant support~\citep{dVetal1997}.  Our new analyses indicate the presence of sign epistasis: we find both negative and positive fitness changes by the addition of single mutations depending on the genetic background.  We also find local fitness maxima and minima in both data sets, leading to severe adaptive constraints.  We then use simulations to compare adaptation of sexual and asexual populations on these empirical rugged fitness landscapes.  We find that populations always get stuck on local peaks, from which they may later escape via the production of genotypes containing multiple mutations.  The simulation results suggest that sex is disadvantageous under most conditions by breaking down escape genotypes.  Only on one landscape and for intermediate recombination rates, sex facilitates the escape from a local to the global peak, hence indicating that this benefit depends on the landscape's topography.  We discuss these findings in the light of experimental evidence for sex benefits during adaptation.

\section{Methods}
\subsection{Strain construction}
\textit{Aspergillus niger} is an asexual filamentous fungus with a predominantly haploid life cycle.  A detailed description of the construction of the \textit{A. niger} strains used in this study is given in~\citep{dVetal1997}. Briefly, haploid segregants were isolated from a diploid between a wild-type strain and a strain containing a single marker mutation on each of its eight chromosomes.  Both the wild-type and the eight-marker strain contained a spore-color marker on its first chromosome (\textit{olvA1} and \textit{fwnA1}, causing olive- and fawn-colored conidiospores, respectively), which was helpful for the isolation of haploid segregants from the diploid mycelium which produced black spores.  The seven marker mutations on the other chromosomes include, in increasing chromosomal order: \textit{argH12} (arginine deficiency), \textit{pyrA5} (pyrimidine deficiency), \textit{leuA1} (leucine deficiency), \textit{pheA1} (phenyl-alanine deficiency), \textit{lysD25} (lysine deficiency), \textit{oliC2} (oligomycin resistance), and \textit{crnB12} (chlorate resistance).  These marker mutations were arbitrarily chosen and the only requirement was their phenotypic detectability.  They were individually introduced using low-dose UV induction, and combined using mitotic recombination to minimize the probability of introducing additional mutations.  Because the wild-type and eight-marker strain had a recent common ancestor, it was unlikely that the segregants differed at more loci than those carrying the marker mutation.  

From the $2^8 = 256$ theoretically possible different haploid segregants, 186 were isolated after forced haploidization on medium containing benomyl from among $\sim$2500 strains tested.  Among those strains, two sets of 32 strains each were present that carry all 32 possible combinations of five markers: the wild type (no mutations), the five strains with a single mutation, the 10 double mutants, the 10 triple mutants, the five quadruple mutants and the single quintuple mutant.  One complete subset, referred to as CS I, involves all combinations of \textit{arg}$^-$, \textit{pyr}$^-$, \textit{leu}$^-$, \textit{oli}$^\textrm{\scriptsize R}$ and \textit{crn}$^\textrm{\scriptsize R}$; CS II contains all possible combinations of \textit{arg}$^-$, \textit{pyr}$^-$, \textit{leu}$^-$, \textit{phe}$^-$ and \textit{oli}$^\textrm{{\scriptsize R}}$.  Hence, these two complete subsets share four of the five marker mutations involved and are not independent.  These complete subsets cover all intermediates between wild-type and quintuple mutant, and thus allow a complete description of the fitness landscape involved~\citep{Wetal2006}.

\subsection{Fitness assay}

Previously~\citep{dVetal1997}, we used the increase in mycelial surface area per unit time on supplemented medium as fitness estimate for these strains, because this measure showed a strong positive correlation with the rate of spore production which is an intuitive measure of fungal fitness~\citep{PT2002}.  There, we used deviations from additivity at the level of log (surface area growth rate) as measure of epistasis.  Here, we use the rate of linear expansion of the radius of a colony spreading from a central source to estimate fitness, because it is believed to be a better estimator of the intrinsic growth rate (i.e. unhindered by competition), and because it is constant in time, which makes departure from additive mutational effects a simple and appropriate way to detect epistasis.  However, radial growth rate and log (surface area growth rate) give almost identical results, because using a linear model, variation in one fitness measure explains $\sim$99.5\% of the variation in the other; the remaining $\sim$0.5\% is explained by the slight curvature of the relationship.  Moreover, many of our analyses require only a rank order of fitness values, and hence are insensitive to these small differences.

	The medium used for the fitness assay is a minimal agar medium supplemented with all amino acids and nitrogen sources required by some of the auxotrophic strains~\citep{dVetal1997}.  The medium was prepared as a single batch and 9-cm petridishes were filled with 20mL medium using a calibrated pump.  Two replicate plates were inoculated in the center with spores from a pre-growth culture of each strain using a platinum needle.  Plates were randomized and incubated at $26^{\circ}\mathrm{C}$ in the dark for 12 days.  Colony diameter was measured after three and 12 days in two perpendicular directions, yielding a single estimate of the radial growth rate per replica plate by dividing the average difference in diameter in the two perpendicular directions at these two time points by twice (to derive radius from diameter) the time elapsed.  All fitness estimates shown are expressed relative to that of the wild type.

\subsection{Simulations}
For the simulations of sexual and asexual populations, we use the Wright-Fisher 
model \citep{F1930,W1931} with a fixed number of haploid individuals $N$. The sequence 
length of genotypes (i.e. number of loci with two alleles each) is denoted by $L$ which 
is 5 for the \textit{A. niger} landscapes, resulting in $2^5=32$ haploid
genotypes.
The algorithm to be employed for the recombination is similar to that described in 
\citep{Kim2005}. 

Let the frequency of the genotype $\sigma$ at generation $t$ be denoted by
$f(\sigma;t)$. At first, the population evolves deterministically in the order
of selection, mutation, and recombination.
By selection and mutation, the frequency of the genotype $\sigma$ at next generation will be

\begin{equation}
p_1\left (\sigma \right ) = \sum_{\sigma^\prime} \mu_{\sigma\sigma^\prime} 
\frac{w\left (\sigma^\prime \right )}{\bar w\left (t \right )} f\left (\sigma^\prime;t\right ),
\label{frequency_pool1}
\end{equation}
where $\bar w(t) \equiv \sum_\sigma f(\sigma;t) w(\sigma)$ is the 
mean fitness of the population at generation $t$ and $\mu_{\sigma\sigma^\prime}$
is the probability that the genotype $\sigma^\prime$ will be the genotype $\sigma$ after
mutation (if no mutation occurs, $\sigma = \sigma'$). 
In what follows, we use a mutation scheme such that 

\begin{equation}
\mu_{\sigma\sigma^\prime} = \left (1-U \right ) \delta_{\sigma\sigma^\prime} + Z_{\sigma\sigma^\prime} 
\frac{U}{L},
\label{mutation_scheme}
\end{equation}
where $\delta_{\sigma\sigma^\prime}$ is the Kronecker delta which takes the 
value 1 if $\sigma =\sigma^\prime $ and 0 otherwise, and $Z_{\sigma\sigma^\prime}$ is 1 
if the Hamming distance between $\sigma$ and $\sigma^\prime$ is 1 and 0 otherwise. This 
mutation scheme implies that a genotype can mutate with probability $U$ and each change occurs 
only at one locus which is chosen at random on the sequence. 

For the recombination scheme, we introduce $W_{\sigma|\sigma'\sigma''}$ 
which is the 
conditional probability that the resulting sequence 
is $\sigma$ in case two sequences $\sigma'$ 
and $\sigma''$ recombine. 
Although we studied three different recombination schemes (free recombination, one
site exchange, and single crossover), we will only present the 
results for free recombination in what follows, because the conclusions for other 
recombination schemes are similar.
Different forms of $W_{\sigma|\sigma'\sigma''}$ for other recombination schemes can be found
in Appendix A.

By free recombination is meant that the new genotype is formed by the random choice of one of 
the alleles at each locus of two genotypes. For example, 
the possible recombinants of two sequences 11101 and 10100 by free recombination 
(with probability $r$) are 10100, 10101, 11100, and 11101, each of which has equal 
probability to contribute to the frequency $p(\sigma)$ in equation~(\ref{frequency_two}).
The free recombination probability can be written as

\begin{equation}
W_{\sigma|\sigma'\sigma''} = \left (1-\delta_{\sigma'\sigma''} \right )
\left [ \left (\delta_{\sigma\sigma'} + \delta_{\sigma\sigma''} \right ) \frac{1-r}{2}
+ \frac{r}{2^{d\left (\sigma',\sigma'' \right )}} + \left (1-\delta_{\sigma\sigma'} \right )
\left (1-\delta_{\sigma\sigma''} \right ) F_{\sigma|\sigma'\sigma''} \frac{r}{2^{d
\left (\sigma',\sigma'' \right )}} \right ] + \delta_{\sigma\sigma'} \delta_{\sigma\sigma''},
\label{free_recombination}
\end{equation}
where $d\left (\sigma',\sigma''\right )$ is the Hamming distance between two sequences in the argument 
and $F_{\sigma|\sigma'\sigma''}$ is 1 if $\sigma$ can be a recombinant of $\sigma'$ 
and $\sigma''$ and 0 otherwise.

Then the frequency of each genotype, after selection, mutation, and recombination, becomes

\begin{equation}
p(\sigma) = \sum_{\sigma',\sigma''} W_{\sigma|\sigma'\sigma''}
p_1(\sigma') p_1(\sigma'') = p_1(\sigma)^2 + 2 \sum_{\langle \sigma',\sigma''\rangle} W_{\sigma|\sigma'\sigma''}
p_1(\sigma') p_1(\sigma''),
\label{frequency_two}
\end{equation}
where $\langle \sigma',\sigma''\rangle$ signifies that the sum runs over all pairs 
of distinct haploid genotypes. 
The actual population at generation $t+1$ is now formed by 
sampling $N$ individual according to the multinomial distribution 
with probability $p(\sigma)$.

\section{Results}
\subsection{No prevailing magnitude epistasis}
\begin{figure}
\epsscale{.95}
\plotone{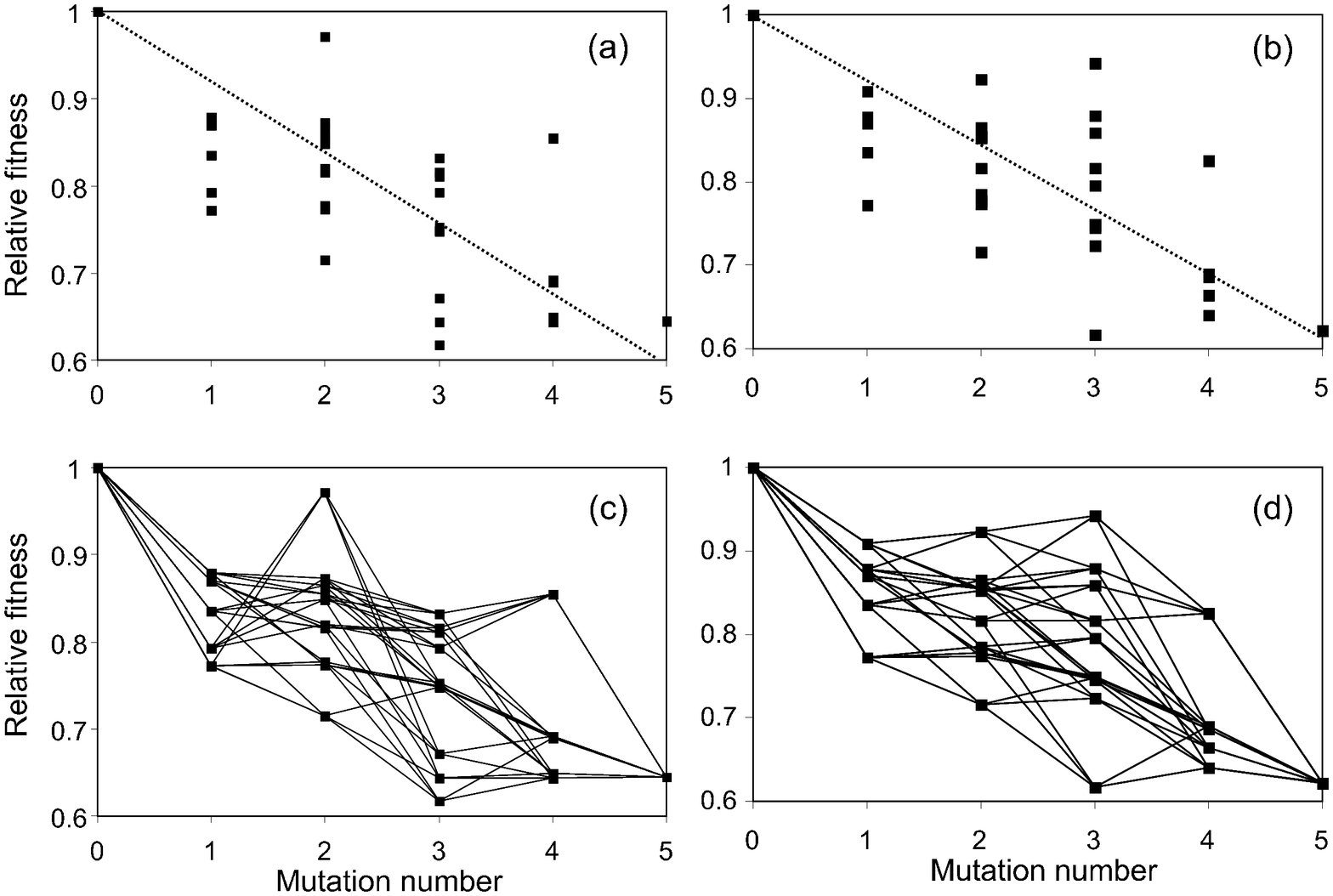}
\caption{\label{Fig:fitness_vs_mutation}Relative fitness versus mutation number for both complete subsets (CS) of 32 strains of \textit{A. niger}; these include \textit{arg}, \textit{pyr}, \textit{leu}, \textit{oli} and \textit{crn} for CS I, and \textit{arg}, \textit{pyr}, \textit{leu}, \textit{phe} and \textit{oli} for CS II.  The overall relationship between fitness and mutation number is best described by a linear model (dashed line), both for CS I (a) and CS II (b).  A model including a quadratic term does not improve the fit significantly (see text), suggesting the absence of a prevailing form of magnitude epistasis.  When the 120 possible (direct, i.e. involving five mutations) pathways between wild-type and quintuple mutant are considered, mutations have negative or positive effect, depending on the genetic background, both for CS I (c) and CS II (d), indicating sign epistasis.  The lines connect genotypes that differ at a single locus (i.e. Hamming distance = 1).}
\end{figure}

Figure~\ref{Fig:fitness_vs_mutation} shows the relationship between relative fitness and mutation number for both complete subsets of \textit{A. niger} strains.  Using surface area growth rate as measure of fitness and regressing log fitness against mutation number, we~\citep{dVetal1997} previously found no evidence for prevailing magnitude epistasis, because adding a quadratic term to the linear regression model did not significantly improve the fit.  Using radial growth rate (RGR) to test for departure from additivity by regressing RGR against mutation number confirms this conclusion.  For both complete subsets of 32 strains, a linear model explains most of the variation in RGR ($r^2 = 0.991$ and 0.992, and the linear coefficient $\alpha = -0.0816$ and $-0.0773$ for CS I and CS II, respectively; see fig. 1a and 1b).  In fact, a linear model explains more variation of the present fitness measure than it did for log fitness in our previous analyses (for which we found $r^2 = 0.888$ and 0.881, for CS I and CS II, respectively), confirming that RGR is likely a more sensitive measure for detecting epistasis as a departure from additivity.  As before, a quadratic term, when added to the model, is estimated to be positive ($\beta$ = 0.0102 and 0.0069 for CS I and CS II, respectively), indicating positive epistasis, but only approaches significance for CS I ($F_{1,30} = 4.135$, $P = 0.0509$) and not for CS II ($F_{1,30} = 1.850$, $P = 0.184$).  Hence, this analysis fails to reveal significant magnitude epistasis in both data sets.

\subsection{Mutational pathways reveal sign epistasis}
Our previous analyses comparing the fitness of double mutants with the average of both single mutants revealed the presence of both negative and positive epistasis, with the majority of combinations showing positive epistasis~\citep{dVetal1997}; this is probably a consequence of the relatively large drop in fitness due to the first mutation, compared to the overall linear trend in figure~\ref{Fig:fitness_vs_mutation}.  We only determined whether the fitness of double mutants was higher (indicating positive epistasis) or lower (indicating negative epistasis) than expected from the fitness of both single mutants, and did not look for sign epistasis.  Here, we seek to reanalyze both complete subsets of 32 \textit{A. niger} strains in order to detect sign epistasis.  To do so, we first analyze the $5! = 120$ possible pathways of five mutations connecting wild-type and quintuple mutant to see whether the addition of some mutations causes fitness to increase rather than decrease.  Figure 1c and 1d shows these mutational pathways for both complete subsets.  Almost 80\% of the 120 pathways exhibit one or more mutational steps causing fitness to increase (95 in CS I and 93 in CS II).  Since all mutations are individually deleterious in the background of the wild-type strain, these fitness reversals indicate sign epistasis among the mutations involved.

To formally demonstrate sign epistasis, we need to show the statistical significance of the observed fitness reversals.  Each complete subset contains 80 unique single-mutation steps, and both subsets combined have 128 unique single-mutation steps (not 160, because four of the five mutations are shared).  Using conventional Bonferroni correction with a cutoff $P$-value of 0.05 for all 128 tests combined renders none of the fitness differences significant.  Linear regression and ANOVA, however, show that genetic differences between these strains are highly significant when tested collectively against the measurement error, and explain almost 95\% of the fitness variation within each complete subset~\cite{dVetal1997}.  The lack of significance of individual tests is therefore due to low statistical power and not to a real absence of fitness differences, and is in part caused by the fact that only two replicate fitness assays were performed (given that both subsets were part of a much greater collection of strains that were assayed).  A possible solution is to use a $P$-value of 0.05 for each individual test and accept that this leads to about six false positives in both data sets combined (i.e. $128 \times 0.05$).  Using this criterion, we find 40 single-mutation steps with a significant fitness effect: 35 declines and five increases.  Although highly unlikely, using this criterion we therefore cannot rule out that all five fitness increases are false positives.

\subsection{The \textit{A. niger} fitness landscapes contain local maxima and minima}
\begin{figure}
\plotone{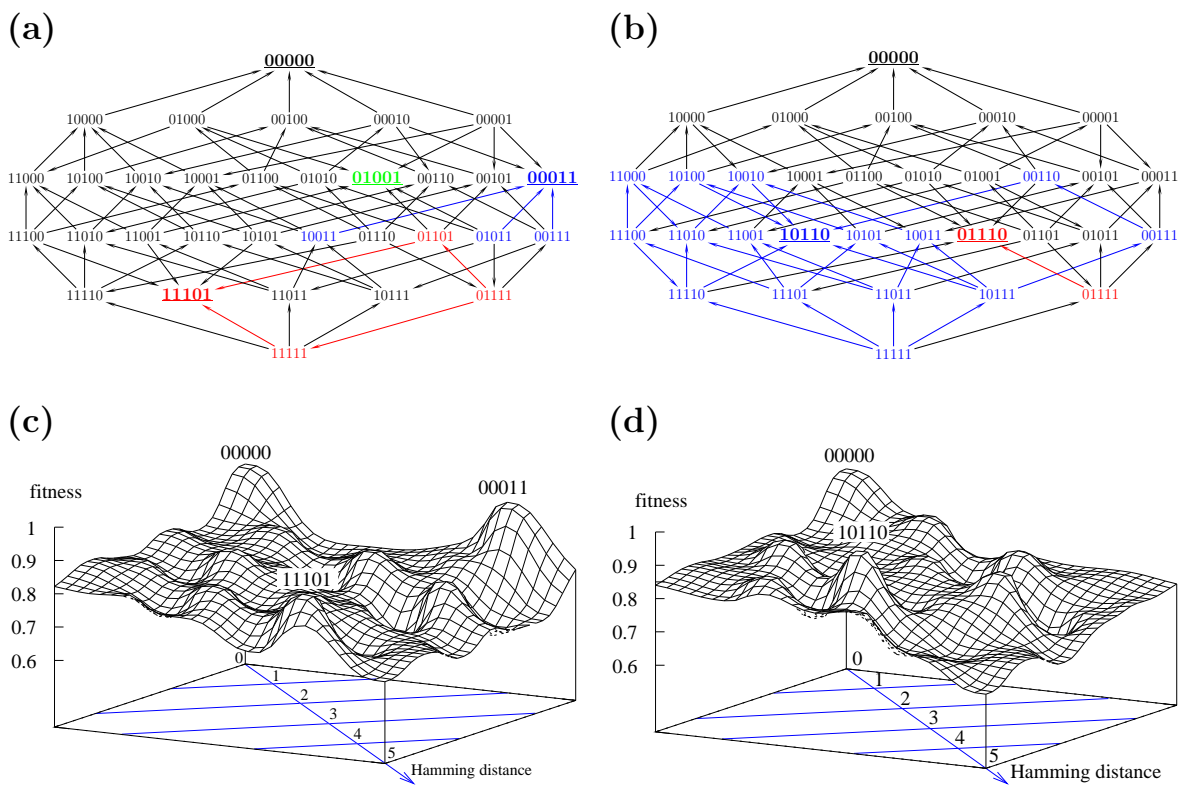}
\caption{\label{Fig:landscapes} Fitness landscapes for both data
  sets. (a) Arrow plot for CS I. Arrows indicate single-mutation steps
  directed towards the genotype with higher fitness. Genotypes
  corresponding to fitness maxima are shown in larger font size and
  underlined. Different colors indicate which genotypes are in the
  basin of attraction of the various fitness maxima; genotype 01001
  shown in green has only itself as basin of attraction. (b) Arrow
  plot for CS II. (c) Rendering of the CS I fitness landscape as a
  two-dimensional surface. The genotypes were arranged in a
  diamond-shaped area which mimicks the arrangement in the arrow plots
  in parts (a) and (b), and the fitness values of the genotypes were
  interpolated by a smooth function. The most prominent fitness maxima
  are highlighted, and the lines in the base plane indicate the positions
  of genotypes with equal numbers of mutations. (d) Two-dimensional
  surface plot for CSII.}
\end{figure}

If sign epistasis is locally consistent, it may lead to local fitness maxima surrounded by genotypes of lower fitness, as well as to local fitness minima surrounded by genotypes of higher fitness.  We analyzed both data sets in order to detect local fitness maxima and minima by comparing the fitness rank of all 32 genotypes with that of their five single-mutation neighbors.  This analysis shows several features that indicate the ruggedness of these fitness landscapes (table~\ref{Table1}, figure~\ref{Fig:landscapes}).  First, the wild-type strain represents the global maximum in both landscapes, but the quintuple mutant does not coincide with the global minimum in either landscape; both data sets contain a genotype with lower fitness.  Second, in total four maxima and three minima are identified in CS I and three maxima and two minima in CS II, emphasizing the ruggedness of these landscapes (see figure 2c and 2d).  To test whether these fitness maxima and minima differ from their mutational neighbors (and are not part of a neutral fitness ``plateau''), we performed one-sample one-tailed $t$-tests with four degrees of freedom to test the fitness difference between the single mean value of the focal genotype and that of its five neighbors.  These showed that three of the four maxima and two of the three minima of CS I are significant, while in CS II two of the three maxima and both minima are significant (see table~\ref{Table1}).  Sequential-Bonferroni correction~\citep{R1989} leaves two maxima (the global and a local one) of CS I significant, as well as the global maximum and global minimum (which is different from the quintuple mutant) of CS II.  Thus, while we are unable to identify individual cases of sign epistasis with statistical confidence, these tests confirm the ruggedness of these landscapes caused by sign epistasis leading to isolated fitness maxima and minima.

\subsection{Asexual adaptation on the A. niger landscapes is constrained}
Next, we consider the consequences of these empirical fitness landscapes for the problem of the evolution of sex by exploring the constraints experienced by sexual and asexual populations adapting on these landscapes.  For this, we make two crucial assumptions.  First, we assume that adaptation happens by the transition (by mutation and/or recombination) from one to another of the 32 genotypes, i.e. by substitutions of wild-type or mutant alleles at the five loci only.  Second, we assume that all fitness differences among strains are real, and neglect the statistical issues involved in their significance.  (If we would interpret non-significant fitness differences as evidence for their neutrality, adaptive evolution on these landscapes would hardly be possible, because not a single of the 120 possible pathways involving the sequential substitution of five single mutations from quintuple mutant to wild type would be accessible in either landscape.)  Instead, we will deal with the (relative) neutrality within these fitness landscapes by considering both sign and magnitude of the fitness differences in our simulations.

We studied the adaptive dynamics of asexual populations for a wide range of parameter 
values. In the strong selection-weak mutation (SSWM) limit ($NU\ln N \ll 1$ and 
$Ns \gg 1$ with $s$  denoting a typical selection coefficient) where clonal interference
\citep{GL1998,PK2007} is unimportant, adaptation of an asexual population can be 
approximated by an adaptive walk on the landscape.  Here we will determine the probability 
that the adaptive walker will arrive at one of the local maxima when it starts from the 
quintuple mutant. To this end, we need to specify the probability that a walker located at 
a sequence $\sigma$ jumps to one of its mutational neighbors, which reads \citep{O2002}

\begin{equation}
P(\sigma\rightarrow \sigma') = \frac{\pi(\sigma';\sigma)}{\displaystyle\sum_{\sigma''} \pi(\sigma'';\sigma)}.
\label{Hopping}
\end{equation}
Here  $\pi(\sigma';\sigma)$ is the fixation probability of genotype
$\sigma'$ introduced as a single copy into a population of genotype
$\sigma$, and the denominator sums these probabilities over the five
nearest (i.e. with Hamming distance 1) neighbors of the genotype
$\sigma$.  In the context of the Wright-Fisher model,
$\pi(\sigma';\sigma)$ can be approximated as $\pi \approx 2s$ when the
selection coefficient $s=w(\sigma')/w(\sigma)-1$  is small and
positive.  We neglect the fixation probability of deleterious
mutations.  In the actual calculation we numerically solve the implicit equation $1-\pi = \exp\left [-(1+s)\pi\right ]$ due to Haldane (1927), which is known to provide an accurate approximation for the exact fixation probability of the Wright-Fisher model~\citep{Betal2006}.  Then we enumerate all fitness increasing paths and assign a probability to each path using the transition probabilities from equation (\ref{Hopping}).  Although the analysis is generally applicable to other initial conditions, we only give the results for the initial condition  $\sigma=(11111)$.

\begin{figure}
\plotone{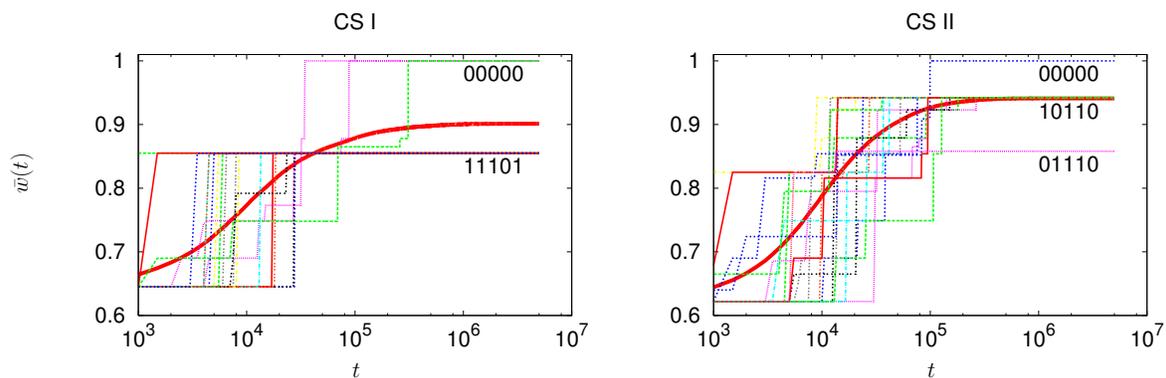}
\caption{\label{Fig:AW} 20 sample adaptive walks starting with genotype 11111 on two empirical landscapes in the SSWM regime ($N = 10^5$ and $U = 10^{-8}$). On the CS I landscape, 17 adaptive walkers arrive at local maximum 11101 and only three walkers could reach the global maximum. The probability to arrive at two other local optima (01001 and 00011) is so low that no walker among 20 samples could reach them (see table~\ref{Table3}). On the CS II landscape, 18 adaptive walks reached local maximum 10110 and only one walker could reach 00000 or the other local maximum 01110. The smooth thick red curves are the average of 10,000 independent adaptive walks.}
\end{figure}

The simulation results in the SSWM regime ($U = 10^{-8}$ and $N = 10^5$) on both landscapes are shown in figure~\ref{Fig:AW} and table~\ref{Table3}.  Clearly, most adaptive walks end at a local maximum, and few reach the global maximum on both landscapes.  The populations stay monomorphic most of the time alternated by brief periods of polymorphism during jumps to another peak.  Mean fitness of 10,000 runs approaches asymptotic values of ~0.90 and ~0.94 for CS I and CS II, respectively.  These asymptotic values are consistent with predictions from the adaptive walk scenario.

Beyond the SSWM regime, there are two conspicuous regimes of asexual
adaptation, which depend on the population size $N$ and mutation rate
$U$~\citep{JK2007}.  At very large $N$, the population behaves as in
the infinite population limit and the evolutionary dynamics are
completely deterministic.  In this regime all possible genotypes are
simultaneously present in the population (but most of them at
extremely small frequencies).  A simple algorithm exists to predict
the trajectories of the most populated genotype that a population will
follow from an arbitrary starting point to the global fitness
maximum~\citep{Jain2005,Jain2007a}, see Appendix B.  For our empirical landscapes these trajectories are:

\begin{eqnarray}
& 11111 \to 11101 \to 00011 \to 00000 & \textrm{CS~I}, \label{Eq:GWI}\\
& 11111 \to 11110 \to 10110 \to 00000 & \textrm{CS~II}.
\label{Eq:GWII}
\end{eqnarray}
Apart from the first step in (\ref{Eq:GWII}), all intermediate genotypes appearing in these trajectories are local fitness maxima.  We find that the deterministic regime is reached in our landscapes for population sizes exceeding  $N\propto 1/U^2$ (see below).  

Between the SSWM and the deterministic regime lies the locally deterministic regime of \citet{JK2007}, where the population behaves deterministically up to a crossover time $T_\times$ when the first local maximum is reached.  This initial stage of adaptation can be described as a  ``greedy'' adaptive walk that always chooses the nearest neighbor genotype of highest fitness and gets stuck at local maxima.  For example, a greedy walker on landscape CS II starting from the sequence 11111 will take the steps $11111 \to 11110 \to 10110$ and on the CS I landscape the walk becomes $11111 \to 11101$.  Beyond the initial stage the behavior is determined by the stochastic escape from local maxima by the creation of multiple mutants~\citep{WC2005}.  Provided the mutation rate is sufficiently small, the time scale for the escape dynamics is well separated from the greedy walk phase.  For larger populations the escape dynamics becomes deterministic as well, taking the form of an ``adaptive flight'' between local maxima, as illustrated in (\ref{Eq:GWI}) and (\ref{Eq:GWII}).   

In the following, the intermediate regime will be referred to as the 'greedy walk' regime.  It sets in when the number $NU$ of mutants produced in one generation exceeds the number $L$ of one-mutant neighbors of a given genotype.  It should however be noted that the boundaries between the different regimes are not sharply defined and may depend on the details of the landscape~\citep{JK2007}.  The greedy walker concept allows us to assign to each local maximum a basin of attraction that contains all genotypes that end up at this maximum under the greedy walk dynamics.  The basins of attraction for both fitness landscapes are illustrated in color in the arrow plots of figure~\ref{Fig:landscapes}.

\subsection{Simulating the effect of sex on the \textit{A. niger} landscape}
Given the adaptive constraints for asexuals, we ask whether sex and recombination can be beneficial for adaptation on these landscapes.  As a general first question, we ask whether recombination among locally optimal genotypes is beneficial in the sense that it will either create the globally most fit genotype (i.e. the wild-type), or a genotype that is part of the basin of attraction of the global optimum.  In CS I, three locally optimal genotypes exist, but recombination among them cannot create the globally optimal wild-type (table~\ref{Table2}).  Similarly, recombination between the two locally optimal genotypes of CS II does not create the high-fitness wild-type.  Rather, recombination has a direct negative effect on offspring fitness on both landscapes.  Although sex cannot generate the global optimum directly, it can produce genotypes with access to the global optima via mutation and selection, i.e. that appear in the basin of attraction of the global optimum (table~\ref{Table2}).  Thus, in a polymorphic population with genotypes occupying local fitness maxima, recombination can release populations that are stuck on local fitness peaks in both empirical landscapes and allow them to evolve towards the global optimum.  However, it is not clear which conditions would lead to a polymorphic population occupying several fitness maxima.

In the following we therefore use simulations to ask whether and when sex provides an adaptive advantage for a population that initially consists of the low-fitness quintuple mutant on these landscapes.  In the SSWM regime, the population remains monomorphic most of the time, and only during the selective sweep there are two segregating sequences (whose Hamming distance is 1 within our mutation scheme).  Hence in this regime recombination cannot play any role and the adaptive dynamics are the same for sexual and asexual populations.  For example, we simulated a sexual population with $r = 1$ and 0.01 using the same parameters as in figure~\ref{Fig:AW}, and we observed the same dynamics as for the asexual population (data not shown).

\subsubsection{Infinite population size}

When the population size is large enough, the population can be polymorphic and recombination may play a role.  Let us first investigate the infinite population limit where the dynamics are deterministic.  The stochastic simulation can then be replaced by simply iterating the mutation-selection recombination equation (\ref{frequency_two}) with equations (\ref{frequency_pool1}), (\ref{mutation_scheme}) and (\ref{free_recombination}) for the population frequencies.  Referring to the asexual trajectories (\ref{Eq:GWI}) and (\ref{Eq:GWII}), we expect that the adaptive dynamics on the CS II landscape are simpler, in that only a single intermediate fitness maximum is involved.  We therefore begin with the analysis of CS II.

\begin{figure}
\plotone{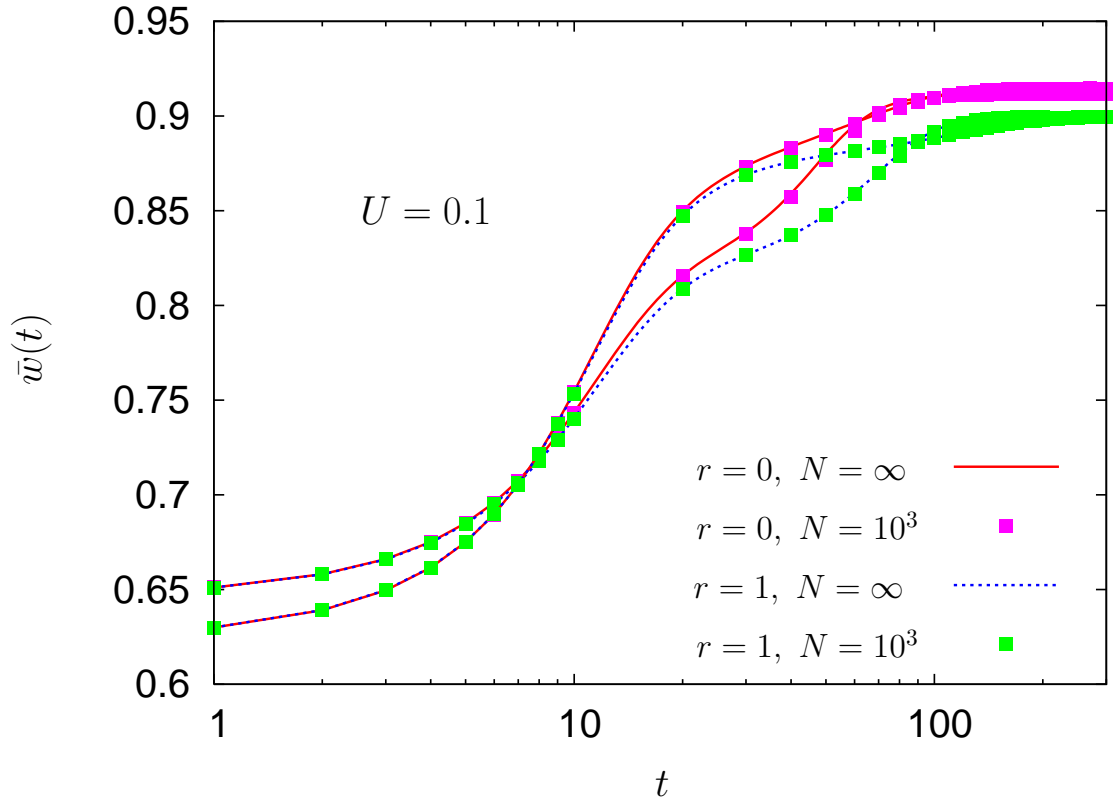}
\caption{\label{Fig:U1e1}The effect of sex during adaptation on
  empirical fitness landscapes for high mutation rates ($U =
  10^{-1}$). Shown are semi-logarithmic plots of mean fitness as a
  function of time (in generations) for 10,000 independent runs with
  ($r = 1$) or without ($r = 0$) recombination. Simulation results for
  finite ($N = 10^3$, symbols) and infinite population size (lines)
  are indistinguishable. The data sets starting at fitness 0.65 are
  results for the CS I landscape, those starting at lower fitness are
  for the CS II landscape. The mutational load (i.e. the difference
  between final asexual fitness and 1) and the recombinational load (i.e. the difference between fitness of the sexual and asexual populations) do not strongly depend on the particular landscape. }
\end{figure}

\begin{figure}
\plotone{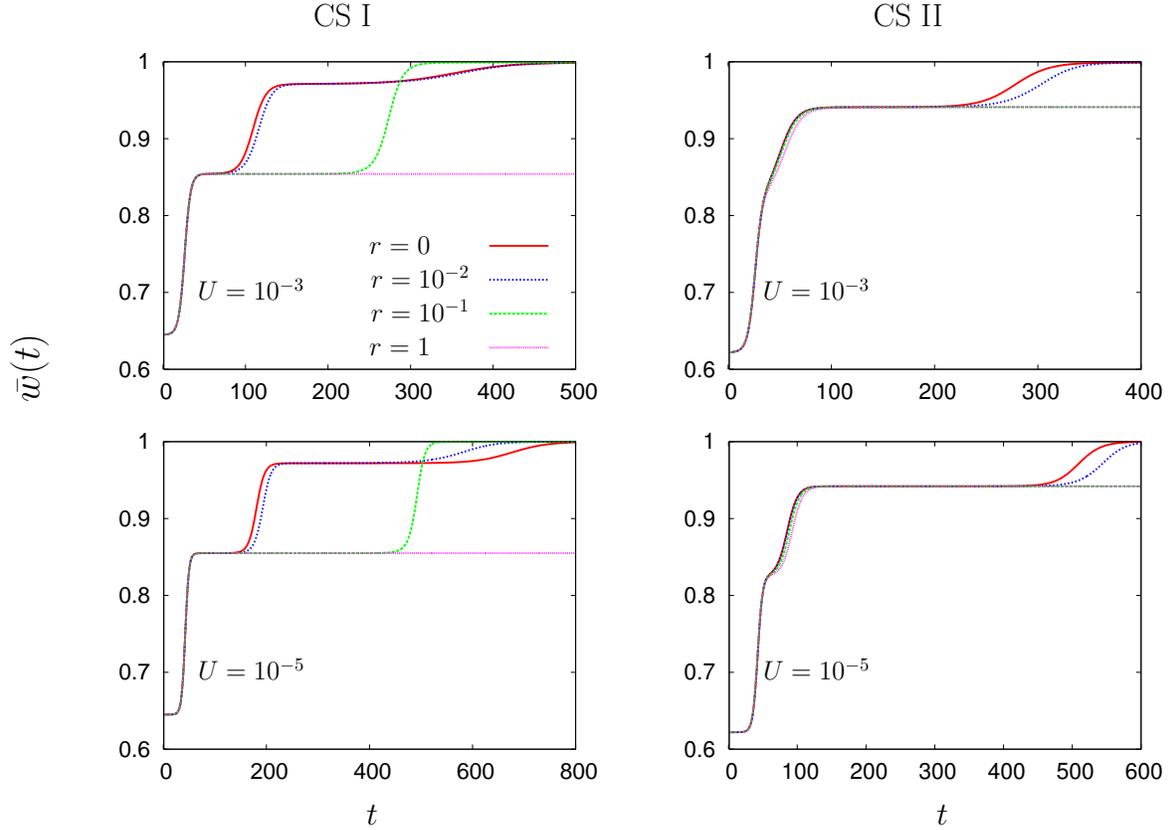}
\caption{\label{Fig:InfCS}Infinite population dynamics during adaptation on two empirical fitness landscapes, CS I (left) and CS II (right), with and without recombination and $U < 10^{-1}$. For CS I, sex is initially deleterious, but low rates of recombination ($r$) later become advantageous during the escape from the first local optimum (genotype 11101 with fitness 0.855), where the second optimum (00011 with fitness 0.972) is bypassed when $r = 10^{-1}$. For CS II, recombination is always disadvantageous.
}
\end{figure}

As figures~\ref{Fig:U1e1} and  \ref{Fig:InfCS} show, recombination on
this landscape never confers an advantage, rather sex is always
deleterious.  The disadvantage of sex is particularly dramatic for $r
\ge  10^{-1}$, where recombination is seen to prevent the escape of
the population from the local maximum 10110.  The suppression of the
escape from a local fitness peak due to recombination is well known
from studies of two-locus models.  In this context the notion of a
\emph{recombination barrier} was introduced, referring to the fact
that the escape rate decreases exponentially with increasing
recombination rate $r$~\citep{S1996,WC2005}. Qualitatively the escape
from the peak is suppressed by breaking up escape genotypes such as 
10000, 00100 and 00010 which are at Hamming distance 2 from the local
peak and at Hamming distance 1 from the wildtype, and which are 
destroyed by recombination with the (dominant) local peak genotype.
For a rough estimate of this effect we may use the formula
\begin{equation}
\label{rcrit}
r_\mathrm{crit} \approx \frac{-s_{\mathrm{ben}}}{1.27 +
  \ln(s_\mathrm{ben})}
\end{equation} 
derived in~\citep{WC2005} within a two-locus model. Equation
(\ref{rcrit}) expresses the critical recombination probability
$r_\mathrm{crit}$ at which suppression of escape sets in as a function of the beneficial selection coefficient $s_\mathrm{ben}$ of the final peak relative to the initial peak.  For the CS II landscape $s_\mathrm{ben} \approx 0.06$, which yields $r_\mathrm{crit} \approx 0.04$, in reasonable agreement with the numerically observed behavior.  In the infinite population case, the escape time may actually diverge at a finite value of $r$, reflecting the appearance of a stationary mutation-selection-recombination equilibrium centered around the initial peak, in addition to the ``true'' equilibrium centered around the global optimum.  Similar \emph{bistable} behavior has been found in deterministic mutation-selection-recombination models with a single peak fitness landscape~\citep{JN2006}.  

The infinite population dynamics on landscape CS I are slightly more complicated.  When $U \ge  10^{-1}$, recombination is always deleterious, because a recombinational load (as well as the mutation load) is observable, similar as for landscape CS II (fig.~\ref{Fig:U1e1}).  However, when $U \le 10^{-3}$, sex appears advantageous for some recombination rates, because the time to arrive at the globally optimal genotype is shorter than for the asexual population, although the time to escape from the local maximum (11101) is longer than for the asexual population (fig.~\ref{Fig:InfCS}).  Closer inspection of this case reveals that recombination allows the population to bypass the intermediate maximum 00011, but this mechanism is operative only in a narrow range (e.g., the range $0.08 \le r \le 0.12$ or $r \le  0.03$ for $U = 10^{-5}$; data not shown).  This implies that the escape time is a nonmonotonic function of $r$ in this regime, with a well-tuned rate of recombination slightly accelerating the adaptation process; similar effects have been observed in two-locus simulations~\citep{WC2005}.  Note that local maximum 11101 is at the same mutational distance ($d = 4$) from both the intermediate maximum 00011 and the global maximum 00000.  The reason for the detour performed by the asexual trajectory (\ref{Eq:GWI}) is that 00011 is closer to the initial point 11111 than the global maximum~\citep{Jain2005}; indeed, an infinite asexual population starting from 11101 moves directly to the global maximum, without visiting 00011 (data not shown).

\subsubsection{Finite population size}
For finite $N$, we simulated 10,000 independent runs for each parameter set.  For $U = 10^{-1}$, the simulation results with $N = 10^3$ are already indistinguishable from the infinite population dynamics for both landscapes (fig.~\ref{Fig:U1e1}).  We also simulated smaller population sizes ($N = 10$, 100) and observed qualitatively similar behavior (data not shown).  We can therefore conclude that recombination is not advantageous on these empirical landscapes for any population size when the mutation rate is high.

\begin{figure}
\plotone{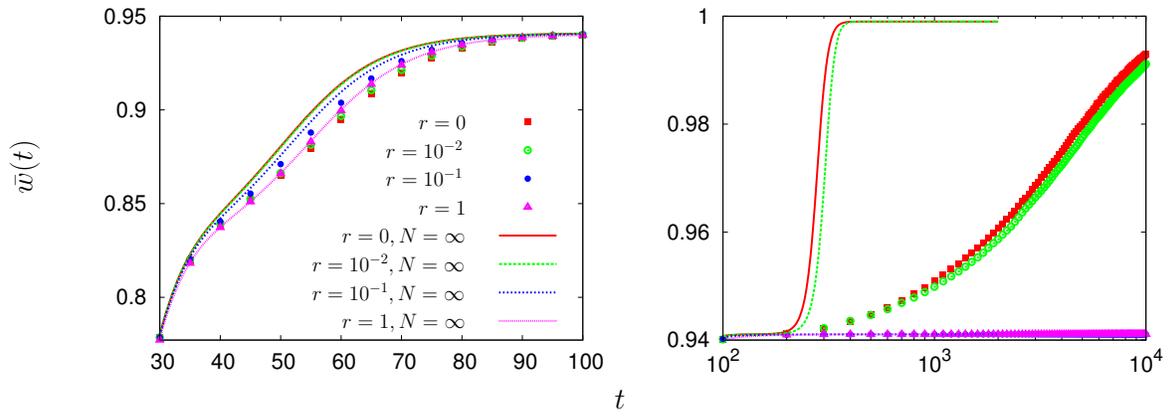}
\caption{\label{Fig:U1e3N1e5}Adaptive dynamics during the greedy walk to (left panel), and escape from (right panel), the local optimum for CS II (genotype 10110 with fitness 0.942). Shown are the trajectories of mean fitness of $10^4$ independent runs for $U = 10^{-3}$ and $N = 10^5$. The dynamics deviate from the deterministic dynamics, and recombination is slightly advantageous during the greedy walk regime, but still disadvantageous during the escape from the local optimum.}
\end{figure}

The situation becomes slightly more complicated at intermediate population size for lower mutation rates.  In the intermediate regime, where the dynamics are indistinguishable from a deterministic greedy walk up to the time at which a local maximum is reached (i.e., up to $T_\times$), sex is generally deleterious on landscape CS II.  This is illustrated by the simulation results for $U = 10^{-3}$ and $N = 10^5$ in figure~\ref{Fig:U1e3N1e5}: there is a small advantage of recombination during the initial greedy walk stage where the dynamics are slightly different from deterministic dynamics, but during the final escape stage recombination is still deleterious.  We actually simulated a wider range of parameters ($N$ up to $10^9$ and $U$ down to $10^{-7}$), but the qualitative behavior is more or less the same as in figure~\ref{Fig:U1e3N1e5}.  Note, however, that this slight advantage of sex during the early stages of adaptation is insignificant for long-term adaptation, which is governed by the time needed to escape from a local peak.

\begin{figure}
\plotone{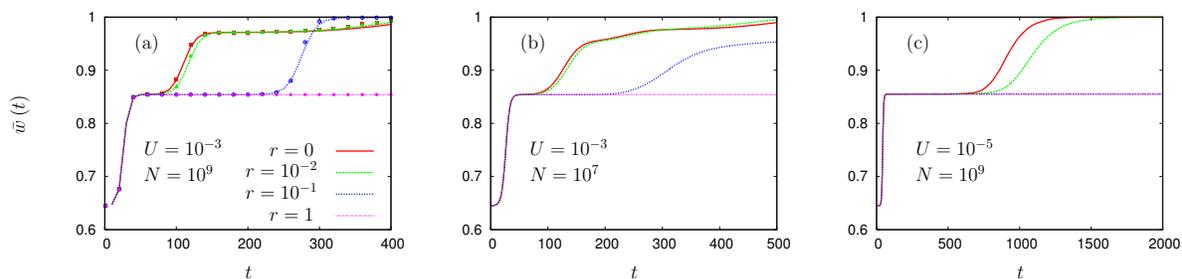}
\caption{\label{Fig:L1land} Adaptive dynamics for CS I for three different parameter sets (mean fitness trajectories of 10,000 independent runs). (a) The adaptive dynamics of the finite populations (symbols) resemble the deterministic behavior (lines). For $r = 10^{-1}$, the population bypasses the second local optimum (00011 with fitness 0.972). Bypassing due to recombination does not happen when either $N$ (b) or $U$ (c) are lower, although there is a slight advantage of sex in the escape stage for $r = 10^{-2}$ for the conditions of panel (b); the lines in panels (b) and (c) show the finite population results.
}
\end{figure}

As expected from the infinite population dynamics, landscape CS I shows more complex adaptive behavior also for finite populations (fig.~\ref{Fig:L1land}).  Let us begin with the initial greedy walk stage.  Since the attractor of the quintuple mutant, i.e. the local maximum 11101, is only one mutation away, we do not expect any difference between the asexual and the sexual greedy walks.  That is, up to the time when fitness reaches the value 0.858, one cannot see any difference in the rate of adaptation between sexual and asexual populations.  Hence, the escape stage will be of main interest for landscape CS I.  For large populations, adaptation is governed by deterministic dynamics (see fig. 7a).  It appears that the escape advantage seen at intermediate $r$ (fig. 7a) depends on a sufficiently large mutation supply rate: when either $N$ (fig. 7b) or $U$ (fig. 7c) is lower, this benefit of recombination disappears.  The fact that we find the escape advantage of recombination on landscape CS I only, indicates that it depends on the topographical details of the fitness landscape.
\section{Discussion}
While epistasis includes all possible forms of deviation from independent gene effects, students of the evolution of sex have almost exclusively been interested in a single particular form of magnitude epistasis, i.e. negative epistasis~\citep{K1993,OL2002,dVE2007,Ketal2007}.  Negative epistasis is one possible source of negative linkage disequilibrium and hence of a long-term advantage of sex by increasing the genetic variation in fitness and accelerating adaptation; alternatively, genetic drift combined with directional selection can cause negative linkage disequilibrium.  Despite increasing efforts to detect negative epistasis in recent years, the support for this form of epistasis is weak~\citep{dVE2007,Ketal2007}.  However, other forms of epistasis may be relevant for the problem of the evolution of sex as well.  For instance, sign epistasis, causing variation in the sign of the fitness effect of mutations across genotypes, may impose adaptive constraints caused by valleys of low fitness separating fitness maxima~\citep{Wetal2005,WC2005}.  Whether and when recombination may facilitate adaptation on such rugged fitness landscapes is largely unknown and depends in part on the topography of the fitness landscape (i.e. the number, height and distance of fitness peaks).

In the present study, we have analyzed fitness data from two sets of 32 strains of the fungus \textit{A. niger} carrying all possible combinations of five mutations with individually deleterious effect.  In a previous study~\citep{dVetal1997}, we used these strains to detect magnitude epistasis by studying the overall relationship between mean fitness and mutation number.  We found no support for prevailing negative or positive epistasis, despite significant magnitude epistasis of both signs for particular pairs of mutations.  Our new analyses were aimed at detecting sign epistasis and describing the fitness landscapes for these sets of five loci.  We found that among the 128 unique single-mutation steps present in the two sets combined, 38 cause fitness increases, despite their individually deleterious effect in the wild type.  In addition, when we compared the fitness of each strain with its five single-mutation neighbors, we found several fitness maxima and minima in both sets of strains, nine of which are statistically significant (four after correcting for multiple comparisons).  These fitness maxima and minima indicate ruggedness of the fitness landscapes resulting from sign epistasis.

We then used simulations with the Wright-Fisher model to study whether and when recombination might facilitate adaptation on these empirical fitness landscapes.  We first found that adaptation was constrained for asexual populations, as only a minority of adaptive walks of finite populations from the low-fitness quintuple mutants ended at the global fitness optimum (see fig.~\ref{Fig:AW}), and even infinitely large populations adapt by sequentially visiting local maxima [equations (\ref{Eq:GWI}) and (\ref{Eq:GWII})].  The simulations showed that adaptation on these rugged landscapes generally involved two distinct stages: the approach to local optimum and the escape from a local optimum to a new optimum.  

Two general conclusions can be drawn from the simulations comparing asexual and sexual populations.  First, recombination provides a small benefit during the approach to a local optimum at intermediate $N$ and $U$, if the local optimum is more than one mutation removed from the ancestral genotype.  This small benefit is probably due to the fact that the alleles involved in the local optimal genotype are in negative linkage disequilibrium due to drift and selection, causing recombination to combine them more often than to disrupt them~\citep{F1930,M1932}.  This effect becomes more pronounced as the distance to the global peak increases.  We have verified in simulations that recombination results in a significant speedup of adaptation in a five-locus system without epistatic interactions (data not shown; see \citet{KK2001} and \citet{Kim2005} for similar results).  However, this small advantage is generally insignificant for long-term adaptation, which is limited by the time needed to escape from a local peak.

Second, during the escape stage the effect of recombination is
generally deleterious and increases the time to escape from the local
optimum.  At high recombination rates ($r = 1$), we only observed
deleterious effects, which are caused by the net break down of escape
genotypes once they arise by mutation~\citep{S1996,WC2005}.  We also
found some conditions where sex and recombination could facilitate the
population's escape.  The fact that we only observed benefits of
recombination for one landscape indicates that the landscape's
specific topography is involved in this benefit.  Also, the escape benefit is only apparent when $N$ and $U$ are sufficiently high (see fig.~\ref{Fig:L1land}), consistent with the dependence of any advantage of recombination on a sufficiently polymorphic population.  In a more general sense, the effect of recombination during the escape stage will depend on conditions where the alleles necessary for the escape genotype are present in negative linkage disequilibrium in genotypes with relatively high fitness.

Few theoretical studies have considered the effect of sex and recombination on rugged fitness landscapes.  \citet{KK2001} found a general disadvantage of sex in a particular two-dimensional landscape with a single `smooth' ridge towards high fitness and no local fitness maxima.  Later, \citet{WW2006} modified this landscape and found conditions where sex does provide an advantage.  The landscapes considered in these papers are similar to ours in that most paths to the global optimum are inaccessible to adaptive evolution, but they differ in that they contain extended neutral networks instead of local fitness peaks.  On the other hand, \citet{HB2003} found a general advantage of recombination in simulations of rather small populations in fitness landscapes generated according to Kauffman's NK-model~\citep{Kauffman1993}.  Together with the work presented here, these fragmentary results suggest that the effects of recombination depend on features of the landscape topography which go beyond the mere presence or absence of sign epistasis.  The precise nature of these features remains to be elucidates in future work.

How do we reconcile our finding of a general lack of sex benefit on
rugged fitness landscapes with the several experimental reports of
faster evolving sexual compared to asexual populations
(e.g. \cite{R2002,dVE2007})?  The reports of sex benefits suggest that
the fitness landscapes involved have different topographies from ours,
e.g. contain smooth areas or ridges allowing unconstrained adaptation.
These differences could have a variety of causes. First, our landscapes are based on the interactions among mutations with individually deleterious effect, which may differ systematically from those involving beneficial mutations.  A similar study of the fitness landscape of TEM-1 $\beta$-lactamase involving five mutations with jointly beneficial effects also found sign epistasis and local ruggedness~\citep{Wetal2006}, but did not find the severe adaptive constraints imposed by isolated local fitness maxima that we found.  We are unaware of other data sets that would allow a more systematic test whether the nature of the mutations studied is responsible for the differences in landscape topographies.  Second, sampling error from studying only a handful of loci may cause our landscape to `miss' smooth ridges or surfaces connecting local peaks, particularly if those are rare.  Third, the topography of fitness landscapes may depend on the level of fitness.  A study of protein fitness landscapes found a relative smooth surface for low-fitness proteins, and ruggedness above a certain fitness value~\citep{Hetal2006}.  If ruggedness appears to be a typical feature of regions of relatively high fitness, our results suggest that sex would be beneficial only for low-fitness individuals, consistent with the negative correlation between fitness and recombination rates observed for many organisms~\citep{HB2003}.

\citet{Setal2007} recently observed a specific advantage for mitotic recombination in the face of sign epistasis in the homothallic fungus \textit{Aspergillus nidulans}.  By comparing adaptation in diploid and haploid strains, they found that four diploid strains that spontaneously reverted to haploidy gained the highest fitness.  Back-crosses of these lines showed that multiple mutations were present, some of which had individually deleterious effect.  Therefore, these lines seemed to have accumulated recessive deleterious mutations when diploid, which showed their combined beneficial effect in haploid recombinants produced in the parasexual cycle.  The advantage of recombination in rugged fitness landscapes may thus depend on the relative length of the diploid and haploid phase of the sexual cycle, which affect the adaptive constraints experienced from fitness valleys.
	
In conclusion, further progress in our understanding of the costs and benefits of recombination will increasingly depend on the combination of experimental studies of the structure of real fitness landscapes with population-genetic simulations comparing different reproductive strategies.  In the present paper we have attempted a modest first step in this direction.

\acknowledgments
We thank Rolf Hoekstra, Duur Aanen, Fons Debets, Yuseob Kim and
Alexander Kl\"ozer for valuable discussion and/or comments on the
manuscript. This work has been supported by DFG within SFB 680
\textit{Molecular Basis of Evolutionary Innovations.}

\appendix
\section{Other recombination schemes}
In this appendix, we describe how two other recombination schemes (one site
exchange and single crossover) are implemented.

One site exchange means that two genotypes may exchange only one locus. Using the same 
exemplar sequences as in the main text (11101 and 10100), the outcome would be 11101, 10100, 10101, and 11100. Unlike 
free recombination, the probability of having one of the above outcomes is different; in 
this example, 11101 and 10100 can occur with conditional probability $\frac{3}{10}$ and 
10101 and 11100 can occur with probability $\frac{1}{5}$ under the 
condition that recombination happens. The explicit form 
of $W_{\sigma|\sigma'\sigma''}$ for the one site exchange rule is

\begin{equation}
W_{\sigma|\sigma'\sigma''} = \left (1-\delta_{\sigma'\sigma''} \right )
\left [ \left (\delta_{\sigma\sigma'} + \delta_{\sigma\sigma''} \right ) \frac{1}{2}
+ \frac{r d\left (\sigma',\sigma''\right )}{2 L} + \left (1-\delta_{\sigma\sigma'} \right )
\left (1-\delta_{\sigma\sigma''} \right ) Z_{\sigma|\sigma'\sigma''} \frac{r}{2 L} \right ] + \delta_{\sigma\sigma'} \delta_{\sigma\sigma''}
\end{equation}
where 

\begin{displaymath}
Z_{\sigma|\sigma'\sigma''} = 
\left \{ 
\begin{array}{cl}
2& \mathrm{ if }~ d\left (\sigma',\sigma''\right )=2 ~\mathrm{ and }~ d\left (\sigma,\sigma'
\right )=d\left (\sigma,\sigma''\right )=1,\\
1& \mathrm{ if }~ d\left (\sigma',\sigma''\right )>2~ \mathrm{ and }~ d\left (\sigma,\sigma'
\right )=1,~d\left (\sigma,\sigma'' \right )=d\left (\sigma',\sigma''\right )-1,\\
1& \mathrm{ if }~ d\left (\sigma',\sigma''\right )>2 ~\mathrm{ and }~ d\left (\sigma,\sigma''\right )=1,~d\left (\sigma,\sigma' \right )=d\left (\sigma',\sigma'' \right )-1,\\
0& \mathrm{ otherwise. }
\end{array}
\right .
\end{displaymath}

A crossover divides each genotype into two parts at the same randomly chosen 
position between two loci and mixes them to form new genotypes. 
If, say, crossover happens between the two parents of our example (11101 and 10100), 
the resulting pair would be either 11100 and  10100 
(with probability $\frac{1}{4}$), or  10101 and 11101 (with probability $\frac{3}{4}$).
As before, only one of the pair will join the next generation.
For the crossover, $W_{\sigma|\sigma'\sigma''}$ takes the form

\begin{equation}
W_{\sigma|\sigma'\sigma''} = (1-\delta_{\sigma'\sigma''})
\left [ \left (\delta_{\sigma\sigma'} + \delta_{\sigma\sigma''} \right ) 
\left ( \frac{1}{2} - \frac{\ell_d - \ell_1}{2 (L-1)} r \right )
+ (1-\delta_{\sigma\sigma'})
(1-\delta_{\sigma\sigma''}) C_{\sigma|\sigma'\sigma''} \frac{r}{2 (L-1)} \right ] + \delta_{\sigma\sigma'} \delta_{\sigma\sigma''},
\end{equation}
where $d$ (although we omit the arguments) is the Hamming distance between 
$\sigma'$ and $\sigma''$, $\ell_i$ ($i=1,\ldots, d$) stands for the location of 
the $i^\mathrm{th}$ different locus between $\sigma'$ and $\sigma''$ 
counted from the left, and
$C_{\sigma|\sigma'\sigma''} = \ell_{j+1} - \ell_{j}$ if the crossover occurring between
loci $\ell_{j+1}$ and $\ell_{j}$ ($j=1,\ldots,d-1$) can result in a genotype $\sigma$
and 0 otherwise.

\section{Adaptive flight trajectories}

Here we briefly outline the procedure that leads to the adaptive
trajectories (\ref{Eq:GWI}) and (\ref{Eq:GWII}) for the most populated
genotype in the infinite population
case; for details see
\citep{Krug2003,Jain2005,Jain2007a}. The key idea is to subdivide
the genotype space into \textit{shells} of equal Hamming distance from
the initial genotype, which in our case is the quintuple mutant
11111. Because all genotypes in a shell are fed by mutations
at the same rate, only the most fit genotype in a shell has a chance
to reach a significant frequency in the infinite population
limit. The largest fitness value within a shell defines the
\textit{shell fitness}. 
Moreover, as the population supply by mutations from the
initial genotype decreases exponentially with increasing Hamming 
distance, distant shells can become highly populated only if
the corresponding shell fitness is larger than the fitnesses 
of all shells that are closer to the initial genotype. In short,
the adaptive trajectory can only contain genotypes that are
\textit{records} in the sequence of shell fitnesses. 

As an example, consider the CSI landscape. The fittest genotypes
in the different shells are 11101 ($d=1$), 11001 ($d=2$), 
00011 ($d=3$), 10000 ($d=4$) and 00000 ($d=5$). As the fitness
of 11001 is lower than that of 11101, and the fitness of 
10000 is lower than that of 00011, the sequence of record
genotypes is $11111 \to 11101 \to 00011 \to 00000$. In general,
some of the record genotypes may disappear from the trajectory
because they are \textit{bypassed} by genotypes of higher fitness that
are located further away from the initial genotype, but for the 
landscapes CSI and CSII this does not occur. A careful analysis
of the deterministic mutation-selection equations shows that
this procdure becomes exact in the limit $U \to 0$
\citep{Jain2007a}.

\clearpage

\begin{deluxetable}{ccrcccrccc} 
\tablecolumns{10} 
\tablewidth{0pc} 
\tablecaption{\label{Table1}Analysis of local and global fitness maxima and minima on both \textit{A. niger} landscapes.}
\tablehead{ 
\colhead{}   &\colhead{}&\colhead{} &  \multicolumn{3}{c}{CS I} &   \colhead{}   & 
\multicolumn{3}{c}{CS II} \\ 
\cline{4-6} \cline{8-10}  
\colhead{Strain} & \colhead{Genotype\tablenotemark{1}}   & \colhead{Neighbors\tablenotemark{2}}    & \colhead{$w$} & 
\colhead{Rank}    & \colhead{Max/Min\tablenotemark{3}}   & \colhead{}    & \colhead{$w$} & \colhead{Rank}&\colhead{Max/Min\tablenotemark{3}}}
\startdata 
1  &00000& 2,3,4,5,6     &1    & 1  & \textbf{\underline{Max}}$^{***}$& & 1     & 1   & \textbf{\underline{Max}}$^{**}$ \\
2  &10000& 1,7,8,9,10    &0.878& 3  &    & & 0.878 & 6   &     \\
3  &01000& 1,7,11,12,13  &0.835& 10 &    & & 0.835 & 13  &     \\
4  &00100& 1,8,11,14,15  &0.870& 5  &    & & 0.870 & 7   &     \\
5  &00010& 1,9,12,14,16  &0.772& 20 &    & & 0.909 & 4   &     \\
6  &00001& 1,10,13,15,16 &0.793& 16 & Min$^{*}$& & 0.772 & 21  &     \\
7  &11000& 2,3,17,18,19  &0.865& 6  &    & & 0.865 & 8   &     \\
8  &10100& 2,4,17,20,21  &0.854& 8  &    & & 0.854 & 11  &     \\
9  &10010& 2,5,18,20,22  &0.773& 19 &    & & 0.923 & 3   &     \\
10 &10001& 2,6,19,21,22  &0.873& 4  &    & & 0.773 & 20  &     \\
11 &01100& 3,4,17,23,24  &0.816& 14 &    & & 0.816 & 16  &     \\
12 &01010& 3,5,18,23,25  &0.716& 24 &    & & 0.852 & 12  &     \\
13 &01001& 3,6,19,24,25  &0.848& 9  & Max& & 0.716 & 26  &     \\
14 &00110& 4,5,20,23,26  &0.778& 18 &    & & 0.855 & 10  &     \\
15 &00101& 4,6,21,24,26  &0.820& 12 &    & & 0.778 & 19  &     \\
16 &00011& 5,6,22,25,26  &0.972& 2  & \textbf{Max}$^{***}$& & 0.785 & 18  &     \\
17 &11100& 7,8,11,27,28  &0.816& 13 &    & & 0.816 & 15  &     \\
18 &11010& 7,9,12,27,29  &0.748& 23 &    & & 0.879 & 5   &     \\
19 &11001& 7,10,13,28,29 &0.832& 11 &    & & 0.748 & 23  &     \\
20 &10110& 8,9,14,27,30  &0.749& 22 &    & & 0.942 & 2   & Max$^*$\\
21 &10101& 8,10,15,28,30 &0.792& 17 &    & & 0.749 & 22  &     \\
22 &10011& 9,10,16,29,30 &0.753& 21 &    & & 0.795 & 17  &     \\
23 &01110& 11,12,14,27,31&0.617& 32 & \underline{Min$^{*}$} & & 0.858 & 9   & Max \\
24 &01101& 11,13,15,28,31&0.810& 15 &    & & 0.617 & 32  & \textbf{\underline{Min}}$^{*}$ \\
25 &01011& 12,13,16,29,31&0.643& 31 & Min& & 0.724 & 25  &     \\
26 &00111& 14,15,16,30,31&0.671& 27 &    & & 0.745 & 24  &     \\
27 &11110& 17,18,20,23,32&0.690& 26 &    & & 0.825 & 14  &     \\
28 &11101& 17,19,21,24,32&0.855& 7  & Max$^*$& & 0.690 & 27  &     \\
29 &11011& 18,19,22,25,32&0.649& 28 &    & & 0.665 & 29  &     \\
30 &10111& 20,21,22,26,32&0.692& 25 &    & & 0.686 & 28  &     \\
31 &01111& 23,24,25,26,32&0.643& 30 &    & & 0.640 & 30  &     \\
32 &11111& 27,28,29,30,31&0.645& 29 &    & & 0.622 & 31  & Min$^*$ \\
\enddata 
\tablenotetext{1}{Zero and 1 indicate the absence or presence of a mutation in chromosomal order: \textit{arg}, \textit{pyr}, \textit{leu}, \textit{oli} and \textit{crn} for the CS I landscape, and \textit{arg}, \textit{pyr}, \textit{leu}, \textit{phe} and \textit{oli} for the CS II landscape.} 
\tablenotetext{2}{Neighbors are genotypes (strain numbers) that differ at a single locus.}
\tablenotetext{3}{``Max'' and ``Min'' indicate the presence of local fitness maxima or minima, with the global maximum and minimum of each landscape underlined; significant fitness maxima and minima tested with 1-sample 1-tailed $t$-tests after sequential-Bonferroni correction indicated in bold face; $^* = P < 0.05$, $^{**} = P < 0.01$, and $^{***} = P < 0.001$.}
\end{deluxetable} 
\clearpage
\begin{deluxetable}{ccrcc} 
\tablecolumns{5} 
\tablewidth{0pc} 
\tablecaption{\label{Table3}Summary of the adaptive walks on both landscapes.}
\tablehead{ 
  \multicolumn{2}{c}{CS I} &   \colhead{}   & 
\multicolumn{2}{c}{CS II} \\ 
\cline{1-2} \cline{4-5} 
\colhead{Maximum} & \colhead{Adaptive walk weight\tablenotemark{1}}&&
\colhead{Maximum} & \colhead{Adaptive walk weight\tablenotemark{1}}}
\startdata 
00000&0.294&&00000&0.134\\
00011&0.033&&10110&0.757\\
11101&0.672&&01110&0.109\\
01001&0.001&&&\\
\enddata 
\tablenotetext{1}{Shown are the probabilities that the adaptive walker which starts at the sequence 11111 will visit one of the fitness maxima that are present (see table~\ref{Table1}).  The wild-type genotype (00000) represents the global maximum in both landscapes, the other are local maxima.} 
\end{deluxetable}
\clearpage 
\begin{deluxetable}{ccc}
\tablecolumns{3} 
\tablewidth{0pc} 
\tablecaption{\label{Table2}Recombination among locally optimal genotypes on both \textit{A. niger} adaptive landscapes.}
\tablehead{ 
\colhead{Parents} & \colhead{Offspring\tablenotemark{1}}   & \colhead{Effect on mean fitness\tablenotemark{2}}}
\startdata 
CS I& & \\
1. 01001 x 00011 & \underline{00001}, 01001, 00011, 01011 &-0.096 \\
2. 01001 x 11101 & 01001, \underline{11001}, 01101, 11101 &-0.015 \\
3. 00011 x 11101 & \underline{00001}, \underline{10001}, 01001, \underline{00101},& \\
                 & 00011, \underline{11001}, \underline{10101}, 10011,&\\
                 & 01101, 01011, 00111, 11101,&\\
                 & \underline{11011}, \underline{10111}, 01111, 11111 & -0.145$^{***}$\\
CS II & & \\
1. 10110 x 01110 & 00110, 10110, 01110, 11110 & -0.030\\
\enddata
\tablenotetext{1}{Genotypes that are part of the basin of attraction of the globally optimal wild-type are underlined (shown in black in fig. 2a and 2b).}
\tablenotetext{2}{Mean offspring fitness -- mean parental fitness, tested with 1-sample 2-tailed $t$-test; $^{***}: P < 0.001$.}
\end{deluxetable}

\clearpage


%
\clearpage

\clearpage

\clearpage

\clearpage

\clearpage

\clearpage

\end{document}